\documentclass[aps,pre,showpacs,showkeys,floatfix,unsortedaddress]{revtex4-1}
\usepackage{amssymb}
\usepackage{amsmath}
\usepackage{graphicx}
\usepackage{amsbsy}
\usepackage{color}
\usepackage{bm}
\usepackage[normalem]{ulem}

\begin{document}
\title{Effective stiffness and formation of secondary structures in a protein-like model}

\author{Tatjana \v{S}krbi\'{c}} 
\email{tatjana.skrbic@unive.it}
\affiliation{Dipartimento di Scienze Molecolari e Nanosistemi, 
Universit\`{a} Ca' Foscari di Venezia,
Campus Scientifico, Edificio Alfa,
via Torino 155,30170 Venezia Mestre, Italy}

\author{Trinh X. Hoang}
\email{hoang@iop.vast.ac.vn}
\affiliation{Center for Computational Physics, Institute of Physics, Vietnam
Academy of Science and Technology, 10 Dao Tan, Ba Dinh, Hanoi, Vietnam}

\author{Achille Giacometti} 
\email{achille.giacometti@unive.it}
\affiliation{Dipartimento di Scienze Molecolari e Nanosistemi, 
Universit\`{a} Ca' Foscari di Venezia,
Campus Scientifico, Edificio Alfa,
via Torino 155, 30170 Venezia Mestre, Italy}

\date{\today}
\begin{abstract}
 We use Wang-Landau and replica exchange techniques to study the effect of an increasing stiffness on the
formation of secondary structures in protein-like systems. Two possible models are considered. In both models, a polymer chain is formed by tethered beads
where non-consecutive backbone beads attract each other via a square-well potential representing the tendency of the chain to fold. In addition,
smaller hard spheres are attached to each non-terminal backbone bead along the direction normal to the chain to mimic the steric hindrance of side chains
in real proteins. The two models, however, differ in the way bending rigidity is enforced.
In the first model, partial overlap between consecutive beads is allowed. This reduces the possible bending 
angle between consecutive bonds thus producing an effective entropic stiffness that competes with a short-range attraction,
and leads to a formation of secondary structures characteristic of proteins. We discuss the low-temperature phase diagram as a function of increasing
interpenetration, and find a transition from a planar, beta-like structure, to helical shape.
In the second model, an energetic stiffness is explicitly introduced by imposing an infinitely large energy penalty for bending above a critical angle between consecutive
bonds, and no penalty below it. The low-temperature phase of this model does not show any sign of protein-like secondary structures.
At intermediate temperatures, however, where the chain is still in the coil conformation but stiffness is significant, we find the two models
to predict a quite similar dependence of the persistence length as a function of the stiffness. This behaviour is rationalized in terms of
a simple geometrical mapping between the two models. Finally, we discuss the effect of shrinking side chains to zero, and find the above
mapping to still hold true.
\end{abstract}


\maketitle
\section{Introduction}
\label{sec:introduction}
Synthetic and biological polymers are always characterized by a certain degree of intrinsic stiffness, and hence fall within
the general class of the so-called semiflexible polymers \cite{deGennes79,Grosberg94,Rubinstein03}.
Sometimes, however, stiffness is neglected to first approximation, often because there exist other constraints concurring to provide an effective stiffness, thus
rendering redundant its explicit inclusion.

In a discrete representation, a polymer can be modelled as a sequence of tethered beads where beads represent monomers, and connectivity between consecutive beads 
represent the effect of covalent chemical bonds holding the polymer together. In the simplest representation, angles between successive chain segments are
uncorrelated, and the polymer is viewed as a Brownian curve, with a spherical symmetric evolution. A slightly more realistic representation includes the excluded
volume constraint accounting for the fact that non-consecutive beads (i.e. monomers) cannot overlap being hard-core objects. This however, does not break the spherical
symmetry of the chain, in the sense that each bead can have any position in space that does not violate both the connectivity and the excluded volume constraints.
For this case, the Edwards continuum model has been traditionally regarded as the paradigmatic description able to provide qualitative prediction of the phase diagram.
A flexible polymer in a good solvent can thus be described by tethered beads model, or by the Edwards model in its continuum version \cite{Doi86}, in which only repulsive interactions, due to
excluded volume, are enforced. The effect of a change in the solvent quality and/or by a decrease in the temperature, can be captured by introducing an additional short-range
attraction between non-consecutive beads (i.e. monomers), so there is a tendency of the polymer to collapse to a globular state that is counterbalanced by the excluded volume
repulsion. At the mean-field level, these competitive interactions can be modelled by the well-known Flory theory of polymers \cite{Bhattacharjee13}. 
As the introduced short-range attraction is still
spherically symmetric, the isotropy condition is preserved even under these more general conditions.

Many polymeric molecules, however, exhibit internal stiffness, as remarked. In this case, the angles between two successive segments are no longer uncorrelated, as in the
case of flexible chains, but display nonvanishing correlations \cite{Ha97,Bhattacharjee13}. For this systems, usually referred to as semi-flexible polymers, the simplest and most popular description
is given by the so-called worm-like-chain (WLC) model, where there is an energy penalty in forming an angle between two successive segments of the chain.
Note that this constraint breaks the spherical symmetry present in flexible polymers. Spatial correlation along the chain decays exponentially, thus introducing a persistence
length, above which the chain is still uncorrelated, and hence flexible, that is directly related to the stiffness parameter.

A typical example of stiff biopolymer is provided by DNA, whose hybridized double strand form has a persistence length typically 100 times larger than that corresponding to
each single strand \cite{Hud05}. This property affects the corresponding phase diagram whose collapsed phase is found to be either a toroid or a rod-like phase, unlike the zero-stiffness
flexible case, that is found to be a compact globule \cite{Bloomfield91,Noguchi98,Stevens01,Hoang14,Hoang15}.

Native compact states in globular proteins are found within a rather different framework \cite{Finkelstein02}. Here, polypeptide chains can still be regarded as bio-polymers displaying a high-temperature 
coil (swollen) phase, and a low-temperature collapsed phase, but with a number of specific features that drastically reduce the number of possible conformations, thus driving to a unique
collapsed target conformation. While the precise mechanism with which this is achieved in real proteins is still to be fully understood, few ingredients are
universally accepted to be part of this mechanism. One is the existence of directional interactions (i.e. hydrogen bonds) that are broken with the solvent and
reformed within the protein backbone, upon collapse. Within the implicit model description used in the present paper, this amounts to breaking the spherical symmetry
of the interactions alluded earlier \cite{Maritan00,Poletto08}.
A second crucial feature making a polypeptide chain very different from a synthetic polymer is given by the presence of side chains that stick out in a plane normal to the
protein backbone. As a matter-of-fact, the
primary structure of a protein is formed by a sequence of aminoacid residues whose structure is always the same apart from the side chains whose specificities can be extracted from an alphabet 
of 20 possible letters \cite{Finkelstein02}. Different side chains give rise to different directional interactions, as well as to different packing constraints as they
must be accommodated upon folding in a way to avoid steric clashes. 

In a previous work \cite{Skrbic16}, we have analyzed the consequences of the inclusion of both constraints on the phase diagram of the folded phase. 
The first constraint was enforced via partial overlapping of consecutive beads, while the second by introducing smaller dummy (i.e. inert hard spheres) spherical beads, mimicking the steric hindrance of side chains, 
at appropriate positions. It was found that a combination of both these ingredients was able to account for the presence of stable secondary structures, such as helices
and beta-like sheets, while each of them, taken singularly, was likely to run into kinetic trapping issues. 

In this paper, we build on this idea and study the detailed reason behind these results. We find that the combination of the two ingredients promotes an effective stiffness
that drives the correct folding, and quantify the dependence of the persistence length as a function of temperature and interpenetration.
Finally, we compare with results stemming from an alternative way of inducing an effective stiffness, and discuss the compatibility of the two models.

The plan of the paper is as follows.

In Section \ref{sec:OPSC_model} we introduce the model and how this induces a bending rigidity. Section \ref{sec:numerical} gives a brief review
of the numerical methods employed, as well as the order parameters used to discriminate between the phases. Sections \ref{sec:effective} and \ref{sec:persistence}
discuss the dependence of the persistence length on both temperature, local interpenetration, and chain lengths. Finally, Section \ref{sec:alternative} reports
a comparison with the model in which there is an infinite energy penalty for large bending and Section \ref{sec:conclusions} draws some conclusions and discusses
open future perspectives.
\section{Inducing effective stiffness in a polymer chain model}
\label{sec:OPSC_model}
\begin{figure}[ht]
\includegraphics[width=5cm]{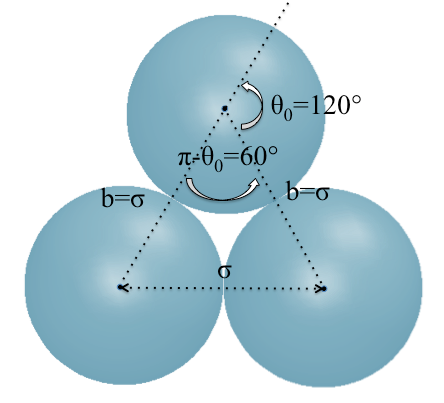}
\includegraphics[width=5cm]{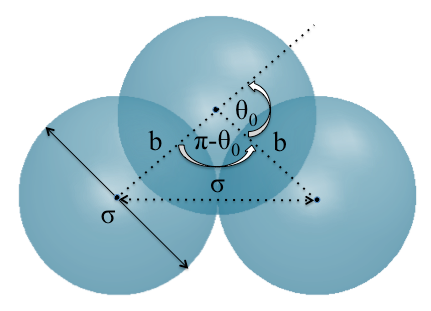} 
\caption{\label{fig:fig1} Effect of partial interpenetration between consecutive monomers for a triplet. (Left) $b/\sigma=1$ so the maximum bond angle
$\theta_0$ is $120^{\circ}$. (Right) $b/\sigma<1$ so the maximum angle is $\theta_0 <120^{\circ}$.
} 
\end{figure}

A common representation of a polymer is provided by a sequence of $N$ beads (monomers), placed at position $\{ \mathbf{r}_{1},
\mathbf{r}_{2},\ldots,\mathbf{r}_N\}$ in space, each of diameter $\sigma$,
connected by a tethering potential holding consecutive beads at a fixed distance $b$ \cite{deGennes79,Grosberg94,Rubinstein03}. Excluded volume
and solvent effects are accounted via an additional potential ensuring that non-consecutive beads cannot interpenetrate,
and attract each other with an energy $\epsilon$ provided that their relative distance is not larger than the interaction range $R_c=\lambda \sigma$,
where $\lambda>1$. 
This additional hard-core square-well (SW) potential can then be written
\begin{equation}
\phi\left(r\right)=\begin{cases}
+\infty\,,\quad \,r < \sigma &\\
-\epsilon,\quad\, \sigma < r < R_c\equiv \lambda \sigma&\\
0, \qquad r > \lambda \sigma&
\end{cases}
\label{eq:sw}
\end{equation}
The model is further defined by imposing the additional constraint that the monomer-monomer
distance $b$ coincides with the bead diameter $\sigma$, so that consecutive beads are tangent to each other ($b/\sigma=1$). In this case, each monomer
can assume any position in space, provided the connectivity constraint to be satisfied. This spherical symmetry combines with the
connectivity constraint to produce a flexible polymer. An effective stiffness can however be induced by allowing partial interpenetration
between consecutive spheres, so that $b/\sigma<1$ \cite{Magee06,Magee07,Bannerman09,Skrbic16}. This breaks the spherical symmetry and gives rise to an effective stiffness as explained
in the cartoon of Figure \ref{fig:fig1}. We will refer to this situation as an \textit{entropic stiffness}.
Let us introduce the bond variable ${\bm \tau}_j=\mathbf{r}_{j}-\mathbf{r}_{j-1}$ so that $\vert{\bm \tau}\vert=b$. A flexible chain is identified by the absence of any correlation between any two bonds
\begin{eqnarray}
  \label{eq:flexible}
  \left \langle {\bm \tau}_{i} \cdot {\bm \tau}_{j} \right \rangle &=& b^2 \delta_{ij}
\end{eqnarray}
where the average $\langle \ldots \rangle$ is over all the possible configurations of the polymer. A more realistic representation of the
polymer is given by the so-called worm-like-chain (WLC) model (also called Kratky-Porod model) \cite{Rubinstein03}, and describes a semi-flexible polymer.
Instead of free joints, the WLC model admits an energetic penalty  every time two consecutive bonds are not parallel.
Denoting by $\theta$ the angle between two consecutive bonds, the WLC model assumes a bending energy proportional to $1-\cos \theta$,
with a proportionality constant $K$ (the stiffness) having dimensions of an energy times a length $L_p$. The latter, can be identified as the
persistent length, the characteristic length appearing in the decay of the tangent-tangent correlation function along the chain (see below).

In discussing the tangent-tangent correlations, it proves convenient to identify the axis of the chain by
a curve coordinate $\mathbf{r}(s)$ parameterized by its arc length $s$, that is the continuum version of $\mathbf{r}_{j}$ and $s_j=jb$ previously
introduced, respectively. Calculations can then be performed in an efficient way by referring to a particular
Frenet frame of unit vectors
$\{\hat{\mathbf{T}}(s),\hat{\mathbf{N}}(s),\hat{\mathbf{B}}(s)\}$
for the tangent, normal and binormal respectively, that are defined as follows
\begin{eqnarray}
\hat{\mathbf{T}}\left(s\right) &=& \frac{\mathbf{r}^{\prime}\left(s\right)}
{\|\mathbf{r}^{\prime}\left(s\right) \|}\\ \nonumber
\hat{\mathbf{N}}\left(s\right) &=& 
\frac{\hat{\mathbf{T}}^{\prime}\left(s\right)}
{\|\hat{\mathbf{T}}^{\prime}\left(s\right)\|} \\ \nonumber
\hat{\mathbf{B}}\left(s\right)&=& \hat{\mathbf{T}}\left(s\right) \times 
\hat{\mathbf{N}}\left(s\right)
\label{eq:coordinates}
\end{eqnarray}
Here primes denote the derivative with respect to the argument.
The Frenet-Serret local coordinates are related by the Frenet-Serret equations \cite{Coexter89,Kamien02} 
\begin{eqnarray}
 \label{eq:Frenet-Serret}
\frac{\partial \widehat{\mathbf{T}}\left(s\right)}{\partial s} &=& \kappa\left(s\right)
\widehat{\mathbf{N}}\left(s\right) \\ \nonumber
\frac{\partial \widehat{\mathbf{N}}\left(s\right)}{\partial s} &=& - \kappa\left(s\right)
\widehat{\mathbf{T}}\left(s\right) + \tau\left(s\right) \widehat{\mathbf{B}}\left(s\right) \\ \nonumber
\frac{\partial \widehat{\mathbf{B}}\left(s\right)}{\partial s} &=& - \tau\left(s\right)
 \widehat{\mathbf{N}}\left(s\right)
\end{eqnarray}
These equations automatically define the curvature $\kappa(s)$ and the torsion $\tau(s)$ as given in
the first and the last equations of (\ref{eq:Frenet-Serret}). In numerical calculations, however, the discrete version of the Frenet-Serret coordinates
and equations will be used.
Denoting 

\begin{eqnarray}
  \label{eq:tangent}
  \widehat{\mathbf{T}}_i=\frac{{\bm \tau}_{i}+{\bm \tau}_{i+1}}{\left \vert {\bm \tau}_{i}+{\bm \tau}_{i+1} \right \vert}
\end{eqnarray}
as the unit vector associated with the bond length ${\bm \tau}_i$,
translational invariance suggests the tangent-tangent correlations $\langle \widehat{\mathbf{T}}_i \cdot \widehat{\mathbf{T}}_j \rangle \equiv 
\langle \widehat{\mathbf{T}}_{\vert i-j \vert} \cdot \widehat{\mathbf{T}} (0) \rangle$ to depend only upon the difference $s_{ij}\equiv \vert s_i-s_j \vert$, so that
\begin{eqnarray}
  \label{eq:persistence}
  \left \langle \widehat{\mathbf{T}}\left(s\right) \cdot \widehat{\mathbf{T}}\left(0\right) \right \rangle &\sim& \exp\left[-\frac{s}{L_p}\right].
\end{eqnarray}
For a flexible chain, the persistence length $L_p \sim b$ as the correlation (\ref{eq:persistence}) drops to zero after a length of the order of a single monomer.
In this case, the maximum angle between two consecutive bonds can be as large as $\theta_0=120^{\circ}$, as illustrated in Figure \ref{fig:fig1} (Left).
However, if partial interpenetration between two consecutive monomers is allowed, so that $b/\sigma<1$, then 
\begin{eqnarray}
\label{eq:angle}
\frac{\theta_0}{2} &=& \arccos\left(\frac{\sigma}{2b}\right)
\end{eqnarray}
thus translating into an effective entropic stiffness whose strength depends upon the ratio $b/\sigma<1$ and ranges between $b/\sigma=1$ (flexible polymer)
to $b/\sigma=1/2$ (infinitely stiff polymer).
As we shall see, this idea is confirmed by explicit calculations.

While the above effect is sufficient to break the spherical symmetry of the chain, simulations carried out in previous work showed
this model to be prone to kinetic trapping effects, in such the system often gets trapped in long living metastable configurations, especially
for long chains.   
In a previous paper \cite{Skrbic16}, we have shown that this problem can be overcome by introducing smaller beads, having
diameter $\sigma_s<\sigma$, and positioned along the chain so to mimic the side chains in real proteins.

For each of non-terminal backbone beads, one defines a normal vector as
\begin{eqnarray}
  \label{eq:normal}
\widehat{\mathbf{N}}_i & = & \frac{{\bm \tau}_{i+1}-{\bm{\tau}_{i}}}
{\vert {\bm \tau}_{i+1}-{\bm \tau}_{i} \vert }
\end{eqnarray} 
where $i=2,\ldots,N-1$. The corresponding binormal vector is then given by
\begin{equation}
\widehat{\mathbf{B}}_i = \widehat{\mathbf{T}}_i \times \widehat{\mathbf{N}}_i .
\end{equation}
Note that $\{\widehat{\mathbf{B}}_i,\widehat{\mathbf{T}}_i,\widehat{\mathbf{N}}_i\}$, ($i=2,\ldots,N-1$) are the discretized version of the Frenet-Serret local coordinates given in Eqs.(\ref{eq:coordinates}) that are frequently used in the continuum approach of the  polymer theory \cite{Kamien02,Poletto08,Bhattacharjee13}.

To each backbone bead a side chain bead is attached in the anti-normal direction with the position
given by
\begin{equation}
\mathbf{r}^{(s)}_i = \mathbf{r}_i - \widehat{\mathbf{N}}_i (\sigma + \sigma_s)/2 .
\end{equation}
The potentials involving side chain beads are just hardcore repulsions that vanish when $\sigma_s \to 0$.

Following the convention set out in our previous paper \cite{Skrbic16}, we shall refer to this model as the overlapping polymer with side chains (OPSC).

\section{Numerical approaches}
\label{sec:numerical}
Our numerical approach hinges on a combination of two different methods, the microcanonical Wang-Landau (WL)\cite{Wang01} and the canonical Replica Exchange \cite{Allen87,Smith02}
Here, we summarize the main points of both methods, referring to our previous work \cite{Skrbic16}, as well as references therein, for additional details.
\subsection{The Wang-Landau (WL) approach}
\label{subsec:WL} 
Aim of the WL approach is to compute the density of states $g(E)$ in the micro-canonical ensemble, from which the whole thermodynamics can be derived in terms of
the canonical partition function
\begin{eqnarray}
\label{eq:partition}
Z\left(T\right)&=&\sum_{E} g\left(E\right)e^{-E/\left(k_B T\right)},
\label{Z}
\end{eqnarray}
where $k_B$ is the Boltzmann constant.

In order to compute $g(E)$ in WL method \cite{Wang01,Taylor09_a,Taylor09_b,Seaton13} we sample polymer conformations according to micro-canonical distribution,
by generating a sequence of chain conformations $A \to B$, and accepting new configuration $B$ with the micro-canonical acceptance probability
\begin{equation}
\label{eq:prob}
P_{acc}(A \rightarrow B)=\min{ \left(1,\frac{w_B \, g(E_A)}{w_A \, g(E_B)} \right)},
\end{equation}
\noindent where $w_A$ and $w_B$ are weight factors ensuring the microscopic reversibility of the moves. 
The set of Monte Carlo (MC) moves, that are accepted or rejected according to the probability (\ref{eq:prob}) includes both local-type moves, such as single-bead crankshaft, reptation 
and end-point, as well as non-local-type moves, as for instance pivot, bond-bridging and back-bite moves, randomly sampled so that on average $N$ beads (or a multiple
of it) are moved to complete a MC step. 

The required density of states $g(E)$ is then constructed iteratively,  by filling suitable energy histograms and controlling their flatness.
We typically assume convergence after 30 levels of iterations, corresponding to a multiplicative factor value of $f=10^{-9}$.

\subsection{The Replica Exchange approach}
\label{subsec:RE} 
Unlike the WL method, the Replica Exchange (also known as Parallel Tempering) method \cite{Swendsen86,Geyer91} is set in the canonical ensemble, where individual  MC runs  carried out
at a fixed temperature \cite{Allen87,Smith02} are computed in parallel and periodically swapped with one another to overcome high energy barriers typical of complex
energy landscapes, as one expects in the present case. A temperature annealing schedule is defined and temperature swapping is allowed only within neighboring temperatures.

Given two replicas, $\Gamma_i$ and $\Gamma_j$  at temperatures $T_i$ and $T_j$ respectively, the swap move leads to a new state, in which $\Gamma_i$ is at $T_j$  and $\Gamma_j$ is at $T_i$ with
acceptance probability given by the detailed balance condition
\begin{eqnarray}
P_\mathrm{swap} &=& \min \left(1,\exp \left[ \left(\frac{1}{k_B T_i} - 
\frac{1}{k_B T_j}\right)(E_i - E_j)\right] \right).
\end{eqnarray}
The choice of replicas to perform an exchange can be arbitrary, but for a pair of temperatures, for which replicas are exchanged, the number of swap move  trials must be large enough to ensure good statistics. The efficiency of a parallel tempering scheme depends on the number of replicas, the set of temperatures chosen to run the simulations, how frequently the swap moves are attempted, and is still a matter of debate. It has been suggested that for the best performance,  the acceptance rate of swap moves must be about 20\%  \cite{Rathore05}.

\subsection{Order parameters}
\label{subsec:order} 
Order parameters are instrumental in identifying and discriminating between different phases. In the present case, a number of different order parameters can be envisaged.

Firstly, we use the specific heat per monomer $C_{V}/(N k_B)$ to pin down the critical temperature $T_c$ of the transition \cite{Skrbic16}. This however does not tell us
what kind of low-temperature phase is obtained, so additional order parameters have been selected in order to highlight them, as elaborated below.

For the helical phase, the simplest order parameter distinguishing it from a coil configuration is the torsion $\tau(s)$, introduced in the Frenet-Serret Eqs.(\ref{eq:Frenet-Serret}) 
An explicit definition of the torsion $\tau(s)$ can be given in terms of the derivative of $\widehat{\mathbf{T}}$ as
\begin{eqnarray}
\label{eq:torsion}
\tau\left(s\right)&=& 
\frac{
\left(\widehat{\mathbf{T}}(s) \times \widehat{\mathbf{T}}^{(1)}(s)\right) \cdot 
\widehat{\mathbf{T}}^{(2)}(s)
}{\left \vert \widehat{\mathbf{T}}^{(1)}(s) \times \widehat{\mathbf{T}}^{(2)}(s)\right \vert^2}
\end{eqnarray}
where we have defined
\begin{eqnarray}
\label{eq:derivative}
\widehat{\mathbf{T}}^{(n)}(s) \equiv \frac{\partial^{n} \widehat{\mathbf{T}}(s)}{\partial s^n}.
\end{eqnarray}
We will be following the prescription given in Ref. \cite{Magee07} to discretize the above derivatives along the lines set in Eq.(\ref{eq:tangent}),
and obtain the probability distribution $p(\tau)$ as a function of the temperature,
as well as the average torsion
\begin{eqnarray}
\label{eq:average}
\overline{\tau}&=& \frac{1}{L}\int_{0}^{L} ds \; \tau\left(s\right).
\end{eqnarray} 
While a coil phase is characterized by a single mode distribution peaked around $\tau=0$, a helix phase is associated with a bimodal distribution of $p(\tau)$ peaked at two
symmetric values, one positive and one negative, corresponding to the two possible enantiomers, right and left helices.

Another order parameter indicating the formation of planar, beta-like structures is given by the average triple scalar product
$\langle \widehat{\mathbf{N}}_{i}\cdot (\widehat{\mathbf{N}}_{j} \times \widehat{\mathbf{N}}_{k})\rangle$ between any triplet of
normal vectors, that vanishes in a planar structure and is non-zero otherwise.

Finally, we will also monitor the average squared radius of gyration, that can be obtained in the WL approach via the density of state $g(E)$ as
\begin{eqnarray}
\label{eq:RgT}
\left \langle R_g^2(T) \right \rangle &=&\sum_{E} \left \langle R_g^2\right \rangle_{E} g\left(E\right) e^{-E/\left(k_B T\right)}.
\end{eqnarray}
and that is large in the coil configuration and small in the globular phase. Hence, it can be reckoned as an order parameter for the globular phase.
In Eq.(\ref{eq:RgT}) $\langle R_g^2 \rangle_E$ is the average radius of gyration at fixed energy $E$.

\section{Entropic stiffness of the OPSC}
\label{sec:effective}
We study chains composed by $N$  monomers (amino-acid residues in protein language), having a size of $\sigma=6$\AA, 
with size of the side chains set at $\sigma_s=5.0$\AA. Using the former as unit of length, we obtain 
a ratio $\sigma_s/\sigma \approx 0.83$, and for the full range of the monomer-monomer distance $1/2 \le b/\sigma \le 1$, the two extreme values representing
the infinitely stiff and flexible polymers. The range of the attractive square-well potential will be set to $R_c=7$\AA $\approx 1.167 \sigma$ in all simulations
\cite{note}.

Different values of $b/\sigma$ and of $N$ will be considered. The former, will affect the effective stiffness of the chain and hence the corresponding
phase diagram, as anticipated. The latter is related to the issue of the finite size affecting the thermodynamic limit.

\begin{figure}[ht]
\includegraphics[width=5cm]{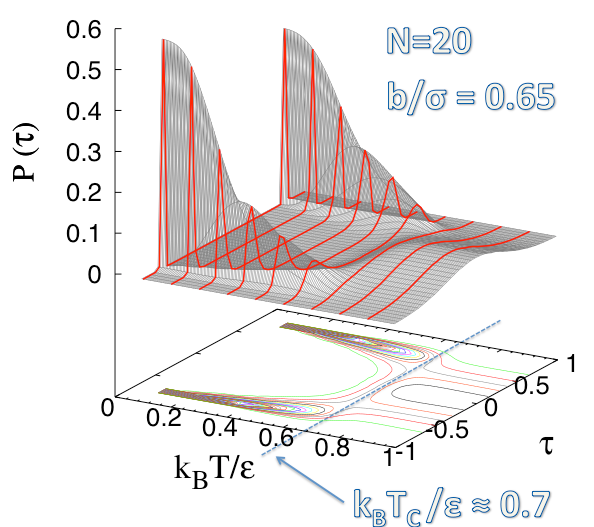}  
\includegraphics[width=5cm]{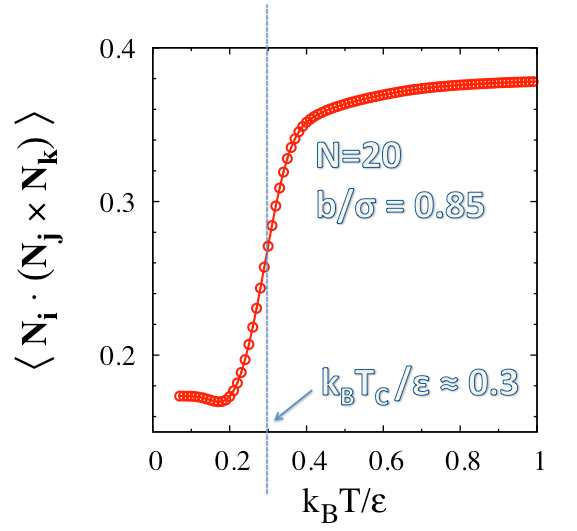}  
\includegraphics[width=5cm]{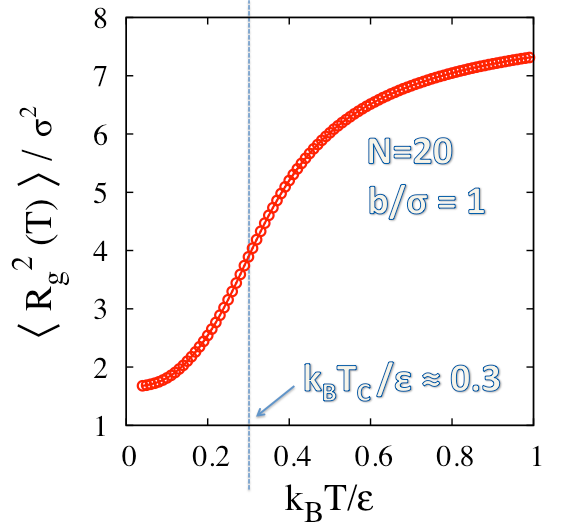}  
\caption{\label{fig:fig2} (Left panel) The torsion probability distribution $p(\tau)$ as a function of the torsion $\tau$ for reduced temperature $k_B T/\epsilon$ above
and below the critical one. 
(Center panel) The average triple scalar product $\langle \widehat{\mathbf{N}}_{i}\cdot (\widehat{\mathbf{N}}_{j} \times \widehat{\mathbf{N}}_{k})\rangle$
 as a function  of the reduced temperature $k_B T/\epsilon$, indicative of the coil-planar structure transition. (Right panel). The average radius of gyration $\langle R_g^2(T) \rangle $
as a function  of the reduced temperature $k_B T/\epsilon$, indicative of the coil-globule transition.
} 
\end{figure}

\begin{figure}[ht]
\includegraphics[width=5cm]{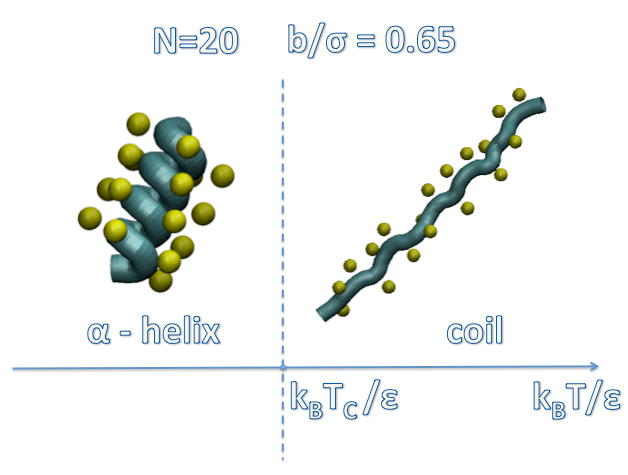}  
\includegraphics[width=5cm]{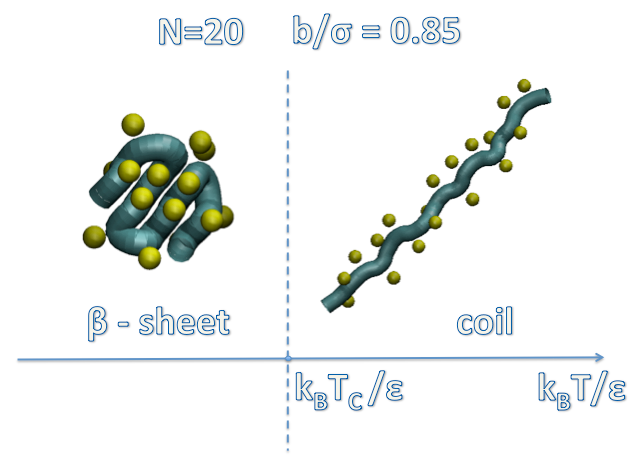}  
\includegraphics[width=5cm]{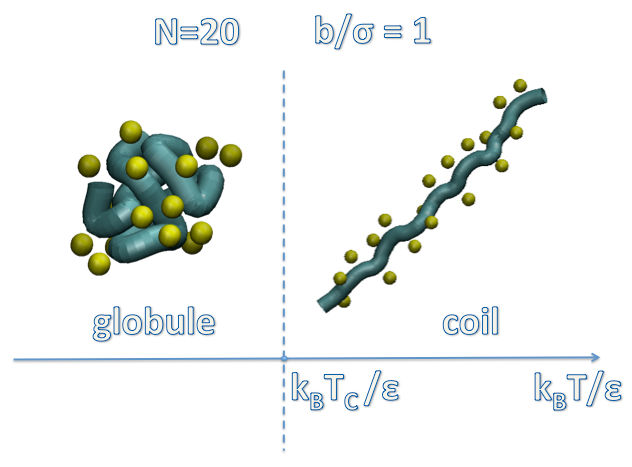}  
\caption{\label{fig:fig3} Phase behaviour of the OPSC model as a function of the temperature and the penetrability ratio. (Right panel) The case flexible $b/\sigma=1$
where the low temperature phase is a globule. (Central panel) The weak penetrability case  $b/\sigma=0.85$ where the low temperature phase is planar-like, 
reminiscent of $\beta-$sheet phases in real proteins. (Left panel) The strong penetrability case $b/\sigma=0.65$ where the low temperature phase
is helix-like, reminiscent of the $\alpha-$helices in real proteins. In all cases, we have used $N=20$ for the number of monomers.
} 
\end{figure}

As one cools down from high temperatures $k_BT/\epsilon \approx 1$ to well below a critical temperature $k_BT_c/\epsilon$, a transition from a swollen (coil) phase
to a low temperature phase is observed as signalled by appropriate order parameter. The actual phase observed depends upon the considered $b/\sigma$,
as illustrated in Figures \ref{fig:fig2} and \ref{fig:fig3} for $N=20$.
Consider the flexible case $b/\sigma=1$ first, whose appropriate order parameter is the radius of gyration $\langle R_g^2(T) \rangle$. This is
reported in right panel of Figure \ref{fig:fig2} where a sudden drop of $\langle R_g^2(T) \rangle$ is observed at $k_B T_c/ \epsilon \approx 0.3$, characteristic
of the collapsed globular phase, as depicted in the corresponding phase diagram of Figure \ref{fig:fig3} (right). Interestingly, here the presence of side
chain appears to be rather ininfluent to the phase behavior, in agreement with the analysis reported in previous work \cite{Skrbic16}.

Upon decreasing the ratio down to $b/\sigma=0.85$, a planar low-temperature phase is observed below $k_B T_c/ \epsilon \approx 0.3$, as signalled by
a drop in the average triple scalar product $\langle \widehat{\mathbf{N}}_{i}\cdot (\widehat{\mathbf{N}}_{j} \times \widehat{\mathbf{N}}_{k})\rangle$ of
Figure \ref{fig:fig2} central panel. The corresponding phase diagram is reported also in the central panel of Figure \ref{fig:fig3}, with
a representative snapshot illustrating the snake-like conformation of the chain roughly confined into a plane. This behavior is reminiscent of the
$\beta$-sheet formation in real proteins and can be reckoned as a direct consequence of the increased stiffness represented by  $b/\sigma<1$ caused by increased
penetrability between consecutive monomers. This, in turn, breaks the spherical symmetry and induces a well defined secondary structure.
For even lower ratio $b/\sigma=0.65$, a coil-helix transition is observed, as identified by the onset of a bimodal distribution symmetrically peaked around $\tau=0$
below a critical temperature  $k_B T_c/ \epsilon \approx 0.7$, as displayed in the left panel of Figure \ref{fig:fig2}. The corresponding phase diagram is then reported
in Figure \ref{fig:fig3} left, where a representative snapshot of the obtained helical phase is also depicted. As in the previous case, this
helical phase is reminiscent of the $\alpha-$helices found in real proteins. While in the $b/\sigma=0.85$ case, the coil-planar structural transition
can be regarded as an effective dimensionality reduction stemming from the symmetry breaking induced by the overlapping of consecutive beads
along the chain, the $b/\sigma=0.65$ case is to be regarded as a chiral breaking from a unimodal to a bimodal torsional distribution, corresponding to the two, a priori equivalent, 
handedness of the helices.

A coil-helix transition can be experimentally measured using various techniques \cite{Cantor80}, and its analysis can be performed by using simple spin models 
\cite{Poland70,Badasyan12,Badasyan10}.
\section{The persistence length}
\label{sec:persistence}
Both the coil-planar and the coil-helix transitions observed in the phase diagrams of Figure \ref{fig:fig3} occur at
a specific temperature, denoted as $T_c$, and are
associated to a thermodynamic discontinuity of the system represented by a sharp peak in the specific heat.
This was explained elsewhere \cite{Skrbic16} and will not be repeated here.

\begin{figure}[ht]
\includegraphics[width=8cm]{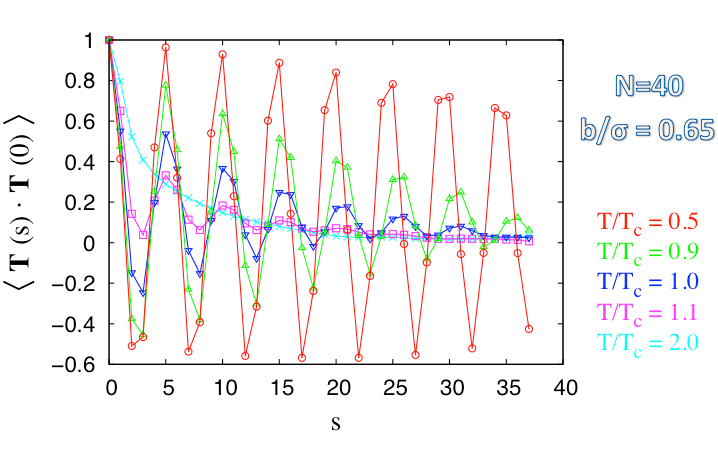}  
\includegraphics[width=8cm]{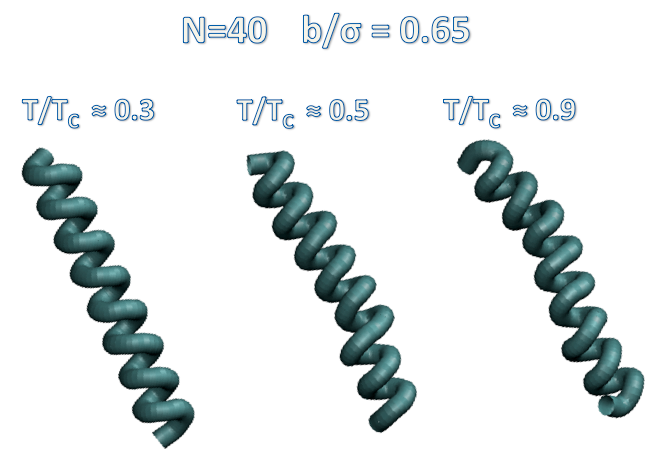}  
\caption{\label{fig:fig4} (Left) The average tangent-tangent correlation function $\langle \widehat{\mathbf{T}}(s) \cdot
  \widehat{\mathbf{T}}(0) \rangle$ as a function of the sequence separation $s$ along the chain, for
  decreasing temperatures across the critical temperature. Here $b/\sigma=0.65$ and $N=40$. (Right) Representative snapshots of
the obtained configurations.
} 
\end{figure}

In the coil-helix transition case, however, an additional interesting feature occurs. Figure \ref{fig:fig4} reports the computed
tangent-tangent correlation function $\langle \widehat{\mathbf{T}}(s) \cdot \widehat{\mathbf{T}}(0) \rangle$
as a function of the sequence separation $s$, as temperature decreases from above $T/T_c>1$ to below $T/T_c<1$ the critical
tempeature $T_c$.
Well above the critical temperature ($T/T_c=2$), the tangent-tangent correlation function exponentially decays, as expected
from Eq.(\ref{eq:persistence}). However, the decay occurs over a length scale significantly longer than $b \approx \sigma$,
as would occur in a flexible chain. Slightly above the critical temperature at $T/T_c \approx 1.1$, the tangent-tangent correlation starts to display oscillatory behaviour
characteristic of a helix, as it will be further discussed below. The period of the oscillation is a measure of the
pitch $P$ of the helix, and remains constant as temperature decreases. Upon cooling below the critical temperature,
the amplitude of the oscillations increases and decays more slowly across the chain, thus indicating an increase in the
effective stiffness of the polymer. Interestingly, oscillations start already at a temperature $T_{FW}>T_c $ indicating
the presence of a structural change before the occurrence of a thermodynamic phase transition.

The oscillatory behavior
can be readily rationalized in terms of an elementary analysis reported in Appendix \ref{app:appa} for an ideal helix.
As detailed in Appendix \ref{app:appa}, the tangent-tangent correlation in a helix clearly oscillates around of an average
value $1/(1+4\pi^2/c^2)$, where $c=P/R$, $R$ and $P$ being the radius and pitch of the helix, respectively.

Note that the amplitude of the oscillation $\Delta F_c$ given by Eq.(\ref{app:eq10}) depends only upon the combination $c$ and not upon the single $R$ and $P$ values.
This feature was already observed in a different, albeit related, context \cite{Snir05,Poletto08}.
However, Figure \ref{fig:figapp} clearly illustrates how the period of oscillation is a direct measue of the helix pitch $P$. In combination
with the amplitude $\Delta F_c$, it provides a measure of both $R$ and $P$ separately.

The moniker of Fisher-Widom (FW) given to temperature $T_{FW}$ stems from the fact that 
this behavior is echoing a similar behavior in liquid theory, usually denoted as a Fisher-Widom line (or point), that can
be ascribed to a structural transition originating from the incipient structural ordering. This is not, however, associated with any discontinuity
in thermodynamics, as it is the transition at $T_{c}$ associated with a peak in the specific heat.  In the present case, the incipient structural ordering is the formation of a particular 
secondary structure (a helix) but this result is clearly more general. 
At high temperatures, the chain is in a coil (swollen) configuration, and consecutive bonds are clearly uncorrelated.
As a result, the tangent-tangent correlation function drops to zero within few units of $\sigma$. As temperature decreases, however, the decay becomes increasingly slower indicating
an increase in the persistence length $L_p$ given by Eq.(\ref{eq:persistence}), as clearly seen in Figure \ref{fig:fig4}. This trend persists until the Fisher-Widom temperature $T_{FW}$ 
is reached, below which an oscillatory
behavior superimposed to the decay is visible, and becomes more and more pronounced as we cross the critical temperature $T_c$.
In the framework of a mean-field theory for a polymer with finite thickness, a similar behavior was observed and ascribed to the disappearing of
a real engenvalue of the corresponding transfer matrix \cite{Marenduzzo05}.
When the system is
in a well characterized helical state, bond vectors display oscillatory behaviour due to the helical path followed by unit vector $\widehat{\mathbf{T}}$
along the chain, as explained in Appendix \ref{app:appa}. Unlike the case of an ideal helix illustrated in Appendix \ref{app:appa}, a 
damping can be observed in the oscillation, so that the helix backbone becomes stiffer and stiffer as we cool down, and the peaks of the oscillation have amplitudes that become
more persistent. By fitting these peaks we can still obtain a value of the persistence length $L_p$. However, care must be exercised in the interpretation of this
result as this should be considered as the ``helix persistence length'' rather than the persistence length of a chain. In the present study, we will neglect this
nuance and use the same definition in both cases.
\begin{figure}[ht]
\includegraphics[width=8cm]{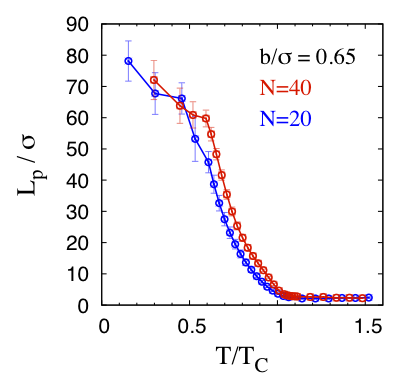}
\includegraphics[width=8cm]{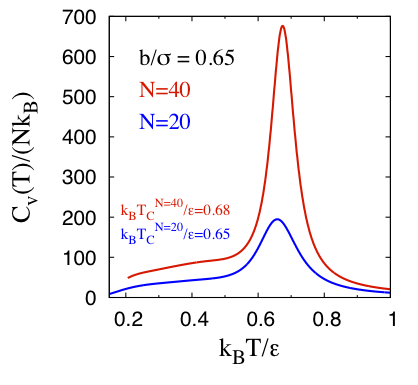}    
\caption{\label{fig:fig5} (Left) Plot of the persistence length $L_p$ as a function of $T/T_c$. (Right) Specific heat per monomer $C_{V}(T)/(N k_B)$
for $N=20$ and $N=40$.
} 
\end{figure}
Figure \ref{fig:fig5} displays the persistence length as a function of the reduced temperature $k_BT/\epsilon$  for a given $b/\sigma=0.65$ and at two different number of beads in the chain ($N=20$ and $N=40$). In both cases, $L_p/\sigma \approx 1$ until
a sudden uprise is observed around $k_BT/\epsilon \approx 0.7$ indicating a progressive increase of the effective stiffness
of the polymer chain. Size effect is visible as the persistence length at a given temperature is larger for $N=40$ than $N=20$.

\section{An alternative model: energetic stiffness}
\label{sec:alternative}
In this section we would like to address the issue of the connection between the OPSC model and another model  where stiffness is enforced in a rather different
way.

We have already discussed the WLC model that is controlled by a bending potential of the form
\begin{eqnarray} 
\label{eq:wlc}
U_{WLC}&=& \frac{K}{2} \sum_{j} \kappa_j^2
\end{eqnarray}
where $K=L_P k_B T$ is the elastic modulus of the bending rigidity (i.e. the stiffness) of the chain, and where we have introduced
\begin{eqnarray}
\label{eq:curvature}
\kappa_j^2 &=& \left \vert \tau_{j+1} - \tau_{j} \right \vert^2=2 b^2\left(1-\widehat{\mathbf{T}}_{j+1} \cdot \widehat{\mathbf{T}}_{j} \right) 
\end{eqnarray}
that determines the local curvature of the chain at the $j$-th monomer position. Note that $\kappa_j$ is the discrete analog of the curvature $\kappa(s)$ appearing in the Frenet-Serret equations
Eq.(\ref{eq:Frenet-Serret}).

In the geometry of chain displayed in Figure \ref{fig:fig1} (left), this is proportional to $1-\cos \theta$ 
with $\theta \le 120^{\circ}$, so that the stiffness is enforced through the bending rigidity introduced through an energetic term (that depends upon temperature), and not through a geometrical constraint.

\begin{figure}[ht]
\includegraphics[width=5cm]{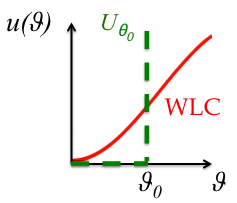}
\caption{\label{fig:fig6} Comparison between the $\theta$ dependence in the WLC model and in the $U_{\theta_0}$ model.}
\end{figure}

An alternative way, used in Refs.\cite{Hoang14,Hoang15}, is in the form of a on-off potential
\begin{eqnarray}
\label{eq:theta0}
U_{\theta_0} &=& \sum_{j} u_j\left(\theta_0\right)
\end{eqnarray}
where
\begin{equation}
u_j\left(\theta_0\right)=\begin{cases}
+\infty\,,\quad \,\widehat{\mathbf{T}}_{j+1} \cdot \widehat{\mathbf{T}}_{j} < \cos \theta_0 &\\
0, \qquad \mbox{otherwise}&
\end{cases}
\label{eq:step}
\end{equation}
Figure \ref{fig:fig6} contrasts this potential with the conventional WLC model, in terms of the $\theta$ dependence.
Essentially, small local directional fluctuations of the chain, within an angle $\theta_0$, are allowed with no energy penalty, but larger ones are prevented.
So, depending upon the actual value of the angle $\theta_0$, the chain may vary from very flexible (when $\theta_0$ is large) to very stiff (when $\theta_0$ is small). We will refer to this as $U_{\theta_0}$ model and note that here $b/\sigma=1$.
Even in the present $U_{\theta_0}$ model, side chains at the same position in space as in the OPSC model are added to make the comparison one-to-one.
Conversely, side chains can be removed both in OPSC model and in the present one, to test for their possible influence on the local stiffness. 
Following the convention set out in our previous work \cite{Skrbic16}, the OPSC in the limit of vanishing side chains will be denoted as overlapping polymer (OP) model.

\begin{figure}[ht]
\includegraphics[width=8cm]{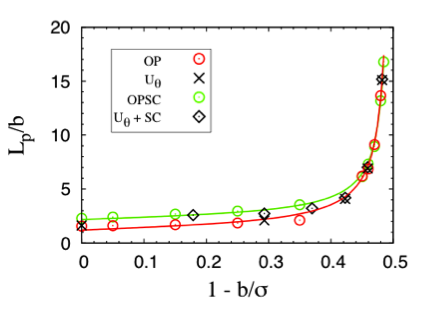}  
\caption{\label{fig:fig7} Comparison of the persistence length dependence $L_p/\sigma$ as a function of $1-b/\sigma$ for the the OP (solid line), OPSC (dashed line), the $U_{\theta_0}$ ($\times$),  
and the $U_{\theta_0}$  model with side-chains ($\Diamond$).
The mapping between the two models has been obtained via Eq.(\ref{eq:angle}) (see Figure \ref{fig:fig1}). 
} 
\end{figure}

At intermediate temperatures, when the chain is in the coil conformation but stiffness is still significant, 
the  $U_{\theta_0}$ model with side chains provides an increasing persistence length $L_p$ as a function of $\theta_0$ akin to the dependence of the OPSC model
for $b/\sigma<1$. This is shown in Figure \ref{fig:fig7}.
Calculations on the $U_{\theta_0}$ model were performed at different values of the critical bending angle $\theta_0$ using replica exchange starting from high
temperature phase, where the chain is in a stiff coil conformation. The range of attraction was set to $R_c=7$\AA as in the OPSC case, and $16$ different temperatures
between $k_B T/\epsilon=1.0$ and $k_B T/\epsilon=0.1$ were used.

Here $N=20$ has been used, and $T>T_c$ in all cases. The tangent-tangent correlations and the corresponding persistence length $L_p$ for the $U_{\theta_0}$ model (with or without side chains) have been computed
following the same steps as in the OPSC model, for decreasing values of $\theta_0$ corresponding to increasing effective stiffness as given by Eq.(\ref{eq:step}).
These values, were then translated in terms of $b/\sigma$ using Eq.(\ref{eq:angle}) and compared with results derived from OPSC model.
No fundamental bias originating from the use of different methodology are expected in this comparison \cite{Skrbic16}.

Figure \ref{fig:fig7} displays the resulting $L_p/b$ as a function of $1-b/\sigma$ for both the OPSC and the $U_{\theta_0}$ model with side chains. 
Both models qualitatively
agree in predicting an increase of the effective bending rigidity, with a remarkable agreement event in quantitative terms.
The lines represent a fit in the form $L_p/b=c_1+c_2 \tan(\pi(1-b/\sigma))$, with $c_1$ and $c_2$ fitting constants.

It is worth stressing that the correct length scale for $L_p$ is $b$ and not $\sigma$, as $L_p$ is measured along the chain. This makes no difference in the 
$U_{\theta_0}$ model, as both quantities coincide, but it does in the OPSC model where $b/\sigma<1$. 

Figure \ref{fig:fig7} also shows the same calculation without the side chains in both models. 
Although there is a slight quantitative difference with the case of side chains,
with this latter being effectively stiffer, the agreement between the two models persists even in this limit.

\begin{figure}[ht]
\includegraphics[width=8cm]{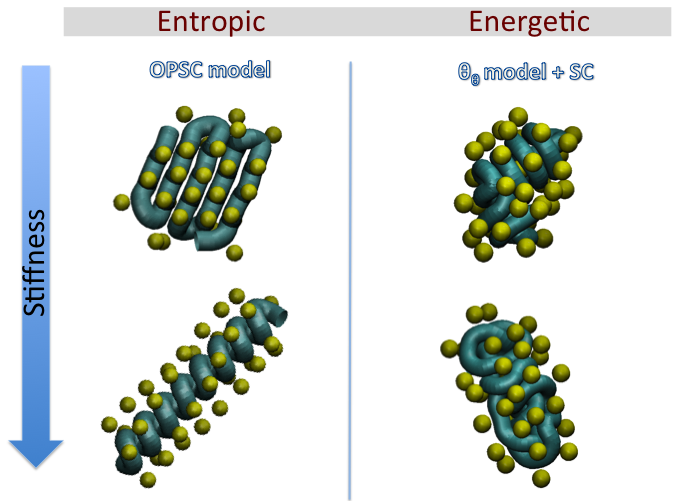}
\includegraphics[width=8cm]{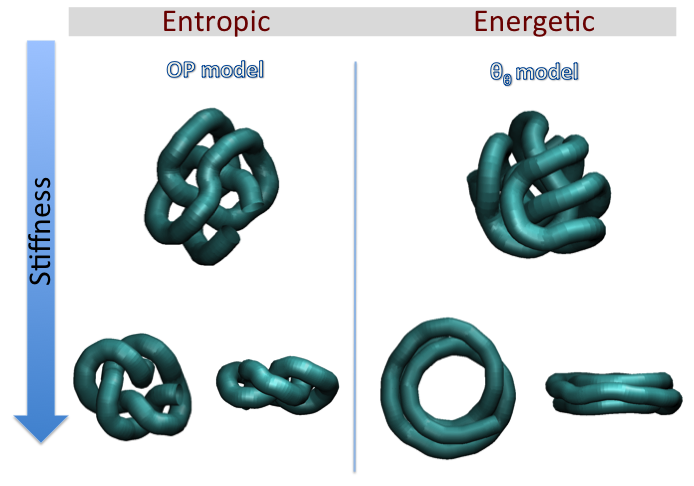}
\caption{\label{fig:fig8} (Left) Ground states obtained with entropic and energetic stiffness, corresponding to the OPSC and the $U_{\theta_0}$ with side chains models,
respectively. Bending rigidity increases from top  ($\theta_0=108^{\circ}$, $b/\sigma\approx 0.85$) 
to bottom ($\theta_0=80^{\circ}$, $b/\sigma\approx 0.65$) . (Right) Same in the case of the OP and $U_{\theta_0}$
models. Here bending rigidity increases from $\theta_0=60^{\circ}$, $b/\sigma\approx 0.58$ top, to
$\theta_0=30^{\circ}$, $b/\sigma\approx 0.52$ bottom. }
\end{figure}

While the intermediate temperatures behavior of the OPSC and the $U_{\theta_0}$ model with side chains are then essentially equivalent in terms of persistence length, their
low-temperatures phases are rather different, as illustrated by the comparison provided in Figure \ref{fig:fig8} (left panel).

For low bending rigidity, corresponding to large  $\theta_0$, one observes globular phases in the $U_{\theta_0}$ model with side chains, 
with no sign of secondary structures formation, contrary to beta-sheet obtained in the OPSC model. Upon increasing
stiffness (i.e. decreasing $\theta_0$), one observes a helical phase in the OPSC model and a hybrid shape in the $U_{\theta_0}$ model with side chains. Here we have used $\theta_0=108^{\circ}$, corresponding to $b/\sigma\approx 0.85$, and  $\theta_0=80^{\circ}$, corresponding to 
$b/\sigma\approx 0.65$, in the more and less flexible cases, respectively.

It is interesting to contrast these findings with the corresponding ones obtained in the absence of side chains. This is depicted in
Figure \ref{fig:fig8} (right panel). Here, we have increased the stiffness significantly to compensate for the absence
of side chains, thus using $\theta_0=60^{\circ}$ (i.e. $b/\sigma\approx 0.58$) in the more flexible case, and 
$\theta_0=30^{\circ}$ ( $b/\sigma\approx 0.52$) in the stiffer case. Here the comparison between the OP and the $U_{\theta_0}$ model,
that give essentially undistinguishable globular phases in the more flexible case ($b/\sigma\approx 0.58$), is striking
in the stiffer ($b/\sigma\approx 0.52$) case. In the  $U_{\theta_0}$ model, a 
 toroidal phase is observed, clearly distinct from a helical phase. This agrees with previous findings within the DNA condensation framework \cite{Hoang14,Hoang15}. By contrast, no such toroidal phase is observed in the OP model with the same stiffness.

\section{Conclusions}
\label{sec:conclusions}
In this paper, we have addressed the problem of the effective stiffness originating in the OPSC model introduced in a recent study \cite{Skrbic16}, as a function
of partial overlap of consecutive beads. The latter represent either monomers, in the polymer language, or amino-acids in protein domain.

As overlap increases from no-overlap ($b/\sigma=1$) to the maximum overlap limit ($b/\sigma=1/2$), the local curvature of the chain is affected
and we have discussed how and why this can be reckoned as an increasing entropic bending rigidity that competes with the short-range attraction to give
different low-temperature phases. In the flexible limit
($b/\sigma=1$) we find a coil-globule transition as expected from conventional synthetic polymers. As this ratio decreases down to $b/\sigma=0.85$ (and hence
stiffness increases), we find
a clear signature of a planar-like low temperature phase, reminiscent of the $\beta-$sheet secondary structure in globular proteins. Upon further decrease down
to $b/\sigma=0.65$ an equally clear signature of a coil-helix transition is found at low temperatures. We argued that the physical origin of this distinction between
these two phases can be traced back to the combined effect of increased entropic stiffness of the local curvature and a change in the excluded volume.

While helical phases were found in past models involving overlapping of consecutive monomers, the present work is the
first, to the best of our knowledge, to show that a certain degree of overlap combined with the effect provided by the side chains, can
lead to a beta-sheet phase.

We have then compared the above findings with those originating from the $U_{\theta_0}$ model with side chains, where the effective bending rigidity has an energetic origin. 
Here the crucial difference from the OPSC model
stems from the fact that there is no interpenetration between consecutive beads, but an explicit infinitely large energy penalty is given for bending of consecutive bonds above a given angle $\theta_0$, with no penalty below it.
With decreasing $\theta_0$, stiffness increases and we find an exact geometrical mapping connecting the $U_{\theta_0}$ model with side chains with the OPSC model.
In full agreement with this mapping, we find that the dependence of the persistence length of the two models in terms of the stiffness to be identical at intermediate temperatures,
where the chain is in the coil but stiff conformation.

  Conversely, at low temperatures, we find the $U_{\theta_0}$ model with side chains to display only globular and hybrid phases for the same values of stiffness.
  For higher bending rigidity and in the absence of side chains, the $U_{\theta_0}$ model displays toroidal phases 
  as it was found in Refs. \cite{Hoang14,Hoang15} 
This means that explicit energetic bending rigidity is \textit{not} compatible with secondary structures typically found in proteins, even in the presence of side chains.
This notwithstanding, it is still noteworthy that the $U_{\theta_0}$ model with side chains and the OPSC share the same phenomenology at intermediate temperatures in terms of an effective
stiff coil conformation, and that this is true whether or not side chains are present.

Interestingly, the OPSC model shares strong similarities with the tube model, introduced by Maritan et al in Ref.\cite{Maritan00} and further reviewed in Refs. \cite{Banavar06,Banavar07}. In that case, the combined
role of entropic bending rigidity and steric effect of the side chains, is played by the thickness of the tube. Indeed, this model was shown
to display \textit{both} beta-like and helical phases upon optimal packing requirement and increasing the range of short-range attractive interactions \cite{Poletto08},
in agreement with what presented here.
The results presented in this paper open up a number of perspectives that we plan to investigate in a future work.
A finite-size dependence study, in terms of increasing the length $N$ and of decreasing the side chain beads size $\sigma_s$, is clearly desirable. A clear-cut comparison with
the tube model in terms of similarities and differences would also be useful \cite{Banavar09}. Another interesting point would be to add a short-range interaction between
the side-chain beads, that was suggested to mimic the effect of hydrogen bonding \cite{Craig06}. More generally, it would be interesting to pursue the effect of additional
ingredients to the OPSC model discussed here, in the strive towards a more realistic description of protein-like systems.
\begin{acknowledgments}
We benefited from discussions with Flavio Romano and Roberto Piazza.
This work was also supported by MIUR PRIN-COFIN2010-2011 (contract 2010LKE4CC). The use of the SCSCF multiprocessor cluster at  the Universit\`{a} Ca' Foscari Venezia is gratefully acknowledged. T.X.H. acknowledges support from Vietnam National Foundation for Science and Technology Development (NAFOSTED) under Grant No. 103.01-2013.16. 
\end{acknowledgments}
\appendix
\section{Tangent-tangent correlation in an ideal helix}
\label{app:appa}
Consider a helix as parametrized by an angular coordinate $0\le \xi \le 2 \pi n_p$, where $n_p$ is the number of different turns (i.e.
pitches). In Cartesian coordinates, its representation with respect to an origin (defined at the origin of the helix) is identified by the vector
\begin{eqnarray}
\label{app:eq1}
\mathbf{r}\left(\xi\right) &=& R \cos \left(\xi\right) \widehat{\mathbf{e}}_{x}+ R \sin \left(\xi\right) \widehat{\mathbf{e}}_{y} + 
  P \frac{\xi}{2\pi} \widehat{\mathbf{e}}_{z}
\end{eqnarray}
Here $R$ and $P$ are the radius and the pitch of the helix, respectively. Then we have
\begin{eqnarray}
\label{app:eq2}
\frac{\partial}{\partial \xi} \mathbf{r}\left(\xi\right) &=& -R \sin \left(\xi\right) \widehat{\mathbf{e}}_{x}+ R \cos \left(\xi\right) \widehat{\mathbf{e}}_{y} + 
  \frac{P}{2\pi} \widehat{\mathbf{e}}_{z}.
\end{eqnarray}

On recalling that the tangent unit vector has the form
\begin{eqnarray}
\label{app:eq3}
\widehat{\mathbf{T}}\left(s\right) &=& \frac{\partial}{\partial s} \mathbf{r}\left(s\right)
\end{eqnarray}
where
\begin{eqnarray}
\label{app:eq4}
\frac{\partial}{\partial s} \mathbf{r}\left(s\right) &=& \frac{1}{s^{\prime}\left(\xi\right)} \frac{\partial}{\partial \xi} \mathbf{r}\left(\xi\right)
\end{eqnarray}
we then have that the condition
\begin{eqnarray}
\label{app:eq5}
\widehat{\mathbf{T}}\left(s\right) \cdot \widehat{\mathbf{T}}\left(s\right)&=& \frac{\partial}{\partial s} \mathbf{r}\left(s\right) \cdot
\frac{\partial}{\partial s} \mathbf{r}\left(s\right)=1
\end{eqnarray}
leads to
\begin{eqnarray}
\label{app:eq6}
s^{{\prime}^2}\left(\xi\right) &=& R^2+\frac{P^2}{4 \pi^2}
\end{eqnarray}
that turns out independent of $\xi$. This expression can then be integrated to get the total length of the string (the backbone of the helix)
\begin{eqnarray}
\label{app:eq7}
L&=& \int_{0}^{2 \pi n_{p}} d \xi \; \sqrt{R^2+\frac{P^2}{4 \pi^2}} = n_{p} \sqrt{P^2+\left(2\pi R\right)^2}.
\end{eqnarray}
Note that this is, in general, different from the ``euclidean length'' $\Lambda= n_{p} P$ that corresponds to the length of the associated
spherocylinder.

\begin{figure}[ht]
  \includegraphics[width=10cm]{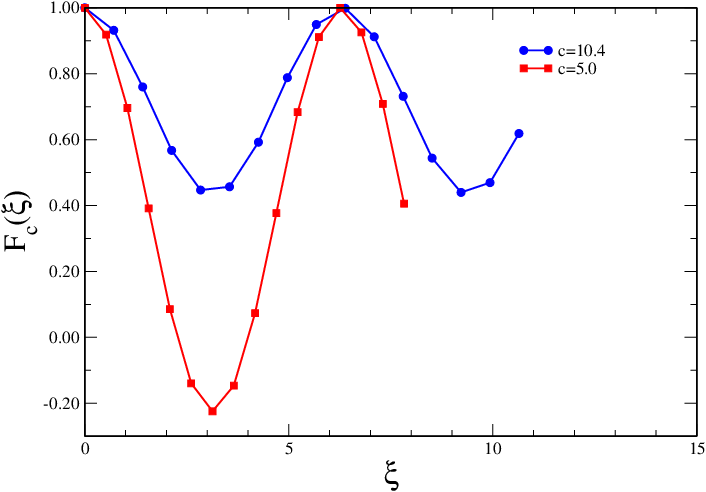}
  \caption{\label{fig:figapp} Tangent-Tangent correlation function $F_c(\xi)$ of an ideal helix as a function of $\xi$ for two different values, $c=10.4$,
  corresponding to $4\pi^2 /c^2\approx 0.39<1$, and $c=5.0$, corresponding to $4\pi^2 /c^2\approx 1.58>1$. 
} 
\end{figure}

From Eqs.(\ref{app:eq1})-(\ref{app:eq4}) we can then compute the tangent-tangent correlation along the helix
\begin{eqnarray}
  \label{app:eq8}
  \widehat{\mathbf{T}}\left(s\right) \cdot \widehat{\mathbf{T}}\left(0\right)&=& F_c\left(\xi\right)
\end{eqnarray}
where
\begin{eqnarray}
  \label{app:eq9}
  F_c\left(\xi\right)&=& \frac{1+\frac{4\pi^2}{c^2}\cos\xi}{1+\frac{4\pi^2 }{c^2}}
\end{eqnarray}
whose behavior as a function of $0\le \xi\le 2\pi$ is reported in Figure \ref{fig:figapp} and it is clearly oscillating. Interestingly, $F_c(\xi)$ depends on the
combination $c=P/R$ and not upon the pitch $P$ and the radius $R$ separately. Figure \ref{fig:figapp} reports the case where $4\pi^2 /c^2\approx 0.39<1$
and hence the oscillation amplitudes are always positive, and the case $4\pi^2 /c^2\approx 1.58>1$ where they can get negative.

Note that the period of the oscillation is identical in the two cases, as it depends only upon the pitch $P$ ($=5$ in the present case), whereas the amplitude of the
oscillation
\begin{eqnarray}
\label{app:eq10}
\Delta F_c \equiv F\left(2 \pi\right)-F_c\left(0\right)=2 \frac{1}{1+\frac{c^2}{4 \pi^2}}
\end{eqnarray}
can be used to measure $c$ and hence the radius $R$.
\vspace{1.0mm} 



\begin{thebibliography}{99}
\bibitem{note} It is worth noticing that an increase of the overlap between consecutive beads, that is a decrease in the value
of $b/\sigma$, effectively translates into an increase of the range of the square-well attraction $R_c$.
\end{thebibliography}
\end{document}